\documentclass{elsart}

\usepackage{amssymb}
\usepackage{graphicx}

\def\BibTeX{{\rm B\kern-.05em{\sc i\kern-.025em b}\kern-.08em
    T\kern-.1667em\lower.7ex\hbox{E}\kern-.125emX}}

\begin{document}

\begin{frontmatter}

\title{Can recurrence networks show small-world property?}

\author[label1]{Rinku Jacob}
\ead{$rinku.jacob.vallanat@gmail.com$}
\author[label1]{K. P. Harikrishnan\corauthref{cor1}}
\ead{$kp_{\_}hk2002@yahoo.co.in$}
\author[label2]{R. Misra}
\ead{$rmisra@iucaa.in$}
\author[label3]{G. Ambika}
\ead{$g.ambika@iiserpune.ac.in$}

\corauth[cor1]{Corresponding author: Address: Department of Physics, The Cochin College,  Cochin-682002, India; Phone No.0484-22224954;  Fax No: 91-22224954.} 

\address[label1]{Department of Physics, The Cochin College, Cochin-682002, India}
\address[label2]{Inter University Centre for Astronomy and Astrophysics, Pune-411007, India}
\address[label3]{Indian Institute of Science Education and Research, Pune-411008, India}

\begin{abstract}
Recurrence networks are complex networks, constructed from time series data, having several practical 
applications. Though their properties when constructed with the threshold value $\epsilon$ 
chosen at or just above the percolation threshold of the network 
are quite well understood, what happens as the threshold increases 
beyond the usual operational window is still not clear from a complex network perspective. 
The present Letter is focused mainly on the network properties at intermediate-to-large 
values of the recurrence threshold, for which no systematic study has been performed so far. 
We argue, with numerical support, that recurrence networks constructed from chaotic attractors 
with $\epsilon$ equal to the usual recurrence threshold or slightly above cannot, in general, show  
small-world property. However, if the  threshold is further increased, the 
recurrence network topology initially changes to a \emph {small-world} structure   
and finally to that of a classical random graph as the threshold approaches the size of the strange attractor. 
\end{abstract}


\end{frontmatter}

\section{Introduction}
In the last two decades, complex network theory has emerged as a popular tool for analyzing complex and 
spatially extended systems \cite {str1,new1,bar1}. It has found applications in a wide range of fields 
including sociology \cite {wat1,was}, communication \cite {new2,law} and biological sciences 
\cite {bul,lon}. The theory of complex networks initially started with random graphs (RG) studied in detail 
by Erd\H os and R\' enyi (E-R) \cite {erd}. For RGs, there is a fixed probability $p$ 
for two nodes being connected and one can show that for a sufficiently large number of nodes $N$, the degree 
distribution $P(k)$  tends to a Poissonian. 

The E-R model guided our thinking about complex networks for many decades until the discovery by Barabasi 
and co-workers \cite {bar2,alb1} nearly two decades back that the topology and structure of most networks 
around us are radically different. For example, many networks in the real world such as, the World Wide 
Web (WWW) \cite {fal}, networks of social interactions \cite {gir}, protein and metabolic networks 
\cite {gui} and technological networks \cite {alb2}, tend to obey certain self-organizing principles in 
their evolution either due to some inherent property of the system or due to the nature of human 
interactions as in social networks. The topology of such networks shows a \emph {scale invariance} with 
the degree distribution obeying a power law, $P(k) \propto k^{-\gamma}$, and the value of $\gamma$ is found  
to be typically between $2$ and $3$. The discovery of such \emph {scale-free} (SF) networks triggered a lot of 
interest in the theory of complex networks. 

Along with the discovery of SF networks, the concept of \emph {small-world} (SW) networks was introduced 
by Watts and Strogatz \cite {wat2}. Though the classical RGs are amenable to a great deal of mathematical 
analysis, they are poor models as far as real networks are concerned. Firstly, they show poor clustering 
and their clustering coefficient (CC) is directly proportional to $p$. Secondly, for a given 
$p$, as the number of nodes $N$ increases, the average degree $<k>$ also increases correspondingly. 
Consequently, for $p$ above a threshold value, the characteristic path length (CPL) of RGs remains very 
small and independent of $N$. 

Watts and Strogatz (W-S) showed that, starting from a ring lattice of $N$ nodes with nearest neighbour 
coupling and randomly re-wiring a small fraction $\beta$ of edges, results in a complex network  with 
high CC compared to the RG and small CPL comparable to that of a RG for a range of 
values of $\beta$. Moreover, for such networks, as $N$ increases, the CPL increases only as $log$ $N$. 
Thus the W-S model displays many characteristic properties of real world networks and provides one 
possibility of obtaining the SW property, often found in real world networks, with different levels of 
complexity by tuning $\beta$.  

The above developments resulted in complex networks and the related measures being applied as tools in 
many areas of research. The most recent among them is the analysis of time series from 
dynamical systems using statistical measures of complex networks. To study many dynamical processes 
in the real world, one often resorts to the analysis of time series obtained from the system. An area of special 
interest is where the underlying system appears to show deterministic nonlinear behavior. The methods and 
measures of nonlinear time series analysis and chaos theory \cite {kan,hil} are commonly used in such cases. 
In the last few years, measures based on complex networks have gained a lot of importance in nonlinear time 
series analysis. All such measures propose a mapping from the time series domain to the network domain and then 
proceed to characterize the dynamical system in terms of the statistical measures of the resulting complex 
network. By doing this, one expects to resolve complimentary features that are not captured by conventional 
methods of time series analysis, especially the structural and topological properties of the 
underlying chaotic attractor. 

Even though several methods have been proposed \cite {zha,lac,mar1} to convert time series into networks, 
an approach incorporating the generic property of \emph {recurrence} \cite {eck} of a dynamical system has 
been prominently applied for the conversion of time series into networks. In this method, the time series is first 
embedded in a suitable dimension $M$ using time delay co-ordinates \cite {gra1} to reconstruct the 
attractor. Every point on the attractor is identified as a node and the network can be constructed in 
two ways, either by taking a fixed number of nearest neighbours \cite {kxu} or by taking a fixed  
hyper-sphere of radius $\epsilon$ with the point as the centre. In this work, we consider the second 
method for the construction of the network where a reference node $\imath$ is connected 
to another node $\jmath$ if the Euclidean distance $d_{ij}$ between the corresponding points on the attractor in the 
reconstructed space is less than or equal to the recurrence threshold $\epsilon$, that is, 
if $d_{ij} \leq \epsilon$. The resulting complex network, called the $\epsilon$ - recurrence network or 
simply recurrence network (RN) \cite {don1,don2}, has been shown to have great potential for a wide range of 
practical applications, from identifying critical transitions in dynamical systems \cite {mar2} to the classification 
of cardio-vascular time series \cite {avi}. Note that, by construction, the RN is an undirected and 
unweighted graph with a symmetric and binary adjacency matrix $\mathcal A$, with elements 
$A_{ij} = 1$ or $0$, depending on whether the two nodes $\imath$ and $\jmath$ are connected or not. 

\begin{figure}
\begin{center}
\includegraphics*[width=16cm]{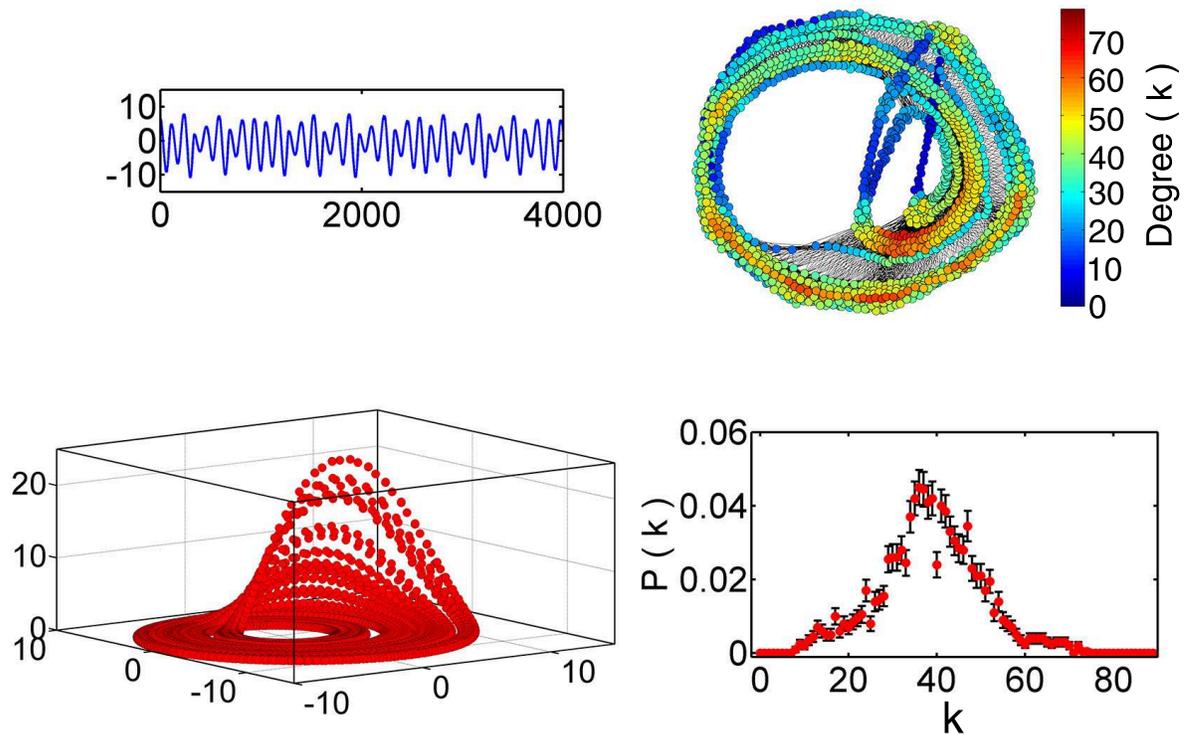}
\end{center}
\caption{Construction of the RN from the time series of the standard R\"ossler 
attractor's $y$ - component using time step $\Delta t = 0.05$ with time delay $\tau = 24$ and $M = 3$.   
The time series and the embedded attractor are shown (top and bottom, respectively) in the left panel while 
the RN and its 
degree distribution are shown in the right panel. The RN is constructed taking every point on the attractor 
as a node and connecting every node to all other nodes within a recurrence threshold of $\epsilon = 0.1$ 
(see text). Error bars of $P(k)$ originate from the fact that for a network with $N$ nodes, the number 
$n$ of nodes with a given degree $k$ has a standard error of $\sqrt n(k)$. 
For $n(k) \rightarrow 0$, its value is normalised to $1$, the minimum count.}
\label{f.1}
\end{figure}

Though RNs and related statistical measures are widely applied in nonlinear time series analysis, their 
properties, as the threshold is increased, are not fully understood from a complex network perspective. 
It is well known that all RNs have 
two properties in common. Firstly, the degree distribution of every RN is unique and is closely related to 
the probability density variations over the embedded attractor from which it is mapped \cite {zou}. 
We will discuss this in detail below. Secondly, there is an absence of long range connections as the maximum  
edge length is limited by the recurrence threshold $\epsilon$. By definition, RNs are random geometric 
graphs (RGG) in the considered system's phase space \cite {zou,don3,dong}. Here RGGs are RGs where each 
vertex is randomly assigned co-ordinates in some metric space according to some prescribed probability 
distribution function, and vertices are connected if and only if 
they are separated by less than a certain maximum distance \cite {dall}. 

In this Letter, we numerically investigate the specific properties of RNs using three primary measures 
of a complex network, the degree distribution, the CC and the CPL. In particular, we consider the 
range of threshold values beyond the small operational window usually used for the construction of 
the RN and check  whether the resulting network can show the properties of either RGs or SF networks or 
behave as a small-world network.  

\section{Numerical Results}
All numerical simulations are done using $N = 2000$ nodes, unless otherwise mentioned. For the construction 
of SF networks, we use the specific scheme of preferential attachment  
discussed in detail in \cite {rava}. Here we start with a small 
number of nodes $m_0$. A new node is then added which gets connected to $m$ number of existing nodes.  
This process is repeated, increasing the number of edges by $m$ for each newly connected node.  
By changing either $m_0$ or $m$ we can construct SF networks with different $\gamma$. 
Here we do both and construct the SF networks using 3 values for $m_0$ (2, 4 and 10) and in each case, 
use 3 different values for $m$, namely, 1, 2 and 4.   
Moreover, the RGs are constructed for different values of $p$. Time series 
from several standard low-dimensional chaotic systems are used for the construction of RNs. For 
continuous systems (flows) in 3D, we use the embedding dimension $M = 3$ and for discrete systems (maps) 
in 2D, we use $M = 2$. For all systems, we use the time series from the $y$ - component and for all 
continuous systems, the time step $\Delta t$ used for numerical integration is $0.05$ with the time 
delay  computed by the automated algorithmic scheme proposed by us \cite {kph1}. The values of time 
delay used for the R\"ossler, Lorenz, Duffing and Ueda systems are $24$, $6$, $25$ and $13$ respectively. 

To get a quantitatively comparable value for the percolation threshold $\epsilon$ and to make the 
comparison between systems possible, we first transform the time series to a uniform deviate so that 
the size of the reconstructed attractor is re-scaled into a unit cube in $M$ dimensions. This transformation 
is not a trivial rescaling as it stretches the embedded attractor uniformly in all directions. We have 
already shown \cite {kph1,kph2} how effective this transformation is in computing conventional measures 
such as, correlation dimension $D_2$ and entropy $K_2$,  
for low as well as high-dimensional systems from time series \cite {kphk11}.  
As a result of the uniform deviate transformation, we have found that it is possible to have a small identical 
range of recurrence threshold $\Delta \epsilon$, that can be taken as the operational window for 
constructing the RN from time series of different systems 
for a given embedding dimension $M$ \cite {jacob}. Here we 
choose the value of $\epsilon$  as the minimum of this range where the giant component 
of the RN just appears, as suggested by Donges et al. \cite {dong}.   
The value of $\epsilon$ is found to be $0.06$ for $M = 2$ and $0.1$ for $M = 3$ for $N \leq 10000$ 
as discussed in detail in \cite {jacob}. The RN from a random (white noise) 
time series is also constructed using $M = 3$ (for comparison with the RNs from chaotic systems) 
whose degree distribution is Poissonian with $<k> \approx 7$ for the selected $\epsilon$.  
We find that this distribution  
coincides with the degree distributions of a RG for $2000$ nodes with $p = 0.0035$ 
and a RGG in $3$ dimensions with the spatial range of connections limited to the same threshold 
$\epsilon = 0.1$  as that of the RN. To draw the connection with the percolation threshold of 
the RN, one can refer to \cite {dong}.  
In Fig.~\ref{f.1}, we show the construction of a typical RN from a standard R\"ossler attractor 
time series. The node degrees of the RN vary over a wide range determined by the local probability density 
variations over the attractor. Specifically, the topology of the RN closely resembles that of the 
embedded attractor, with nodes corresponding to regions of high probability density in the attractor 
having higher degree in the RN and vice versa. This is evident from  Fig.~\ref{f.1}. 

\begin{figure}
\begin{center}
\includegraphics*[width=16cm]{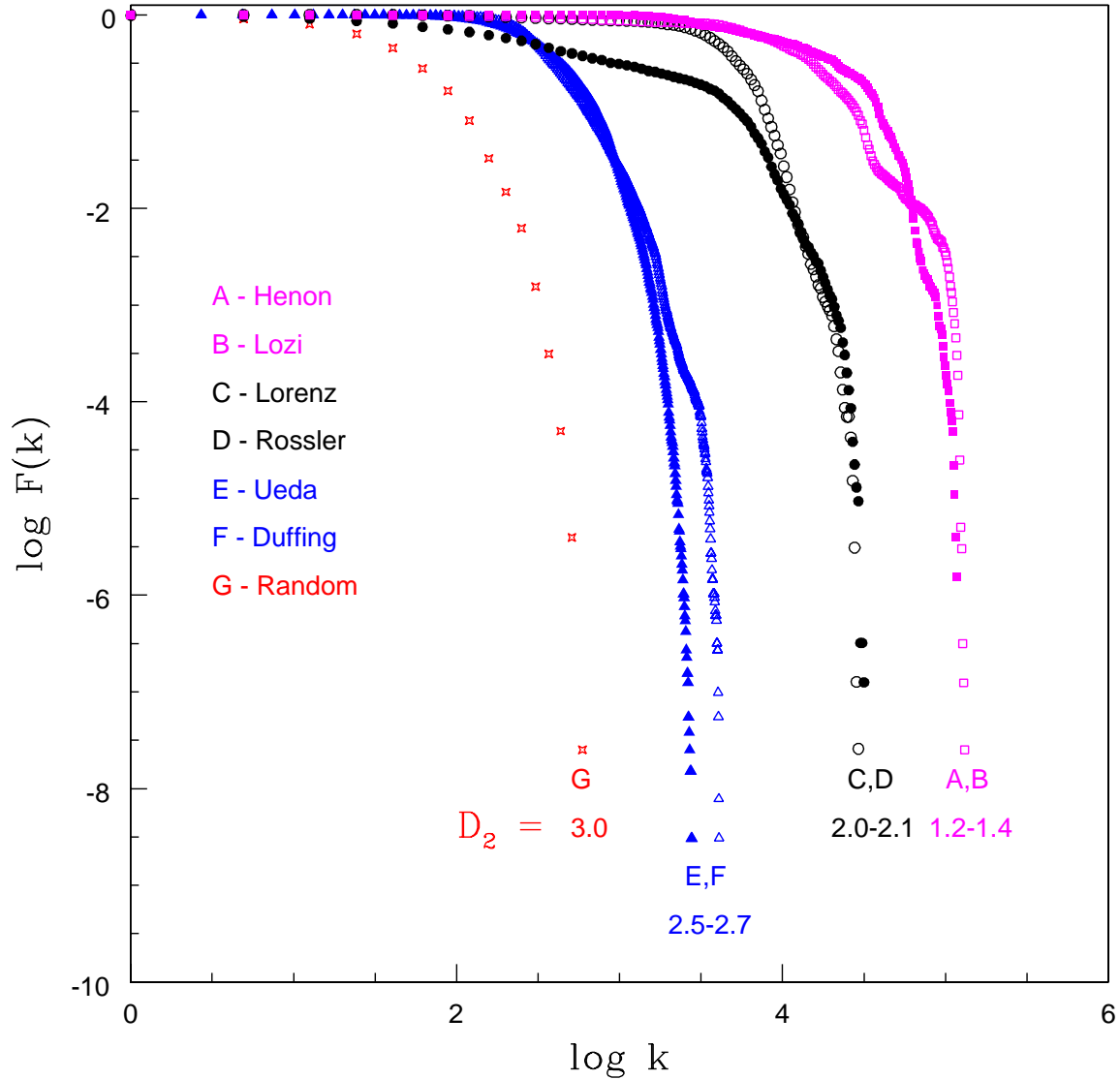}
\end{center}
\caption{Cumulative degree distribution of RNs from several low-dimensional 
chaotic attractors and a random time series. The chaotic attractors are the H\' enon (open square), 
Lozi (solid square), Lorenz (open circle), R\"ossler (solid circle), Duffing (open triangle) and 
Ueda (solid triangle) systems whose $D_2$ ranges are also indicated. The parameters used for all the 
systems are  given in \cite {spr}. We observe that the position where $F(k)$ decreases sharply 
follows a pattern in accordance with the dimension of the attractor.}
\label{f.2}
\end{figure}

\begin{figure}
\begin{center}
\includegraphics*[width=16cm]{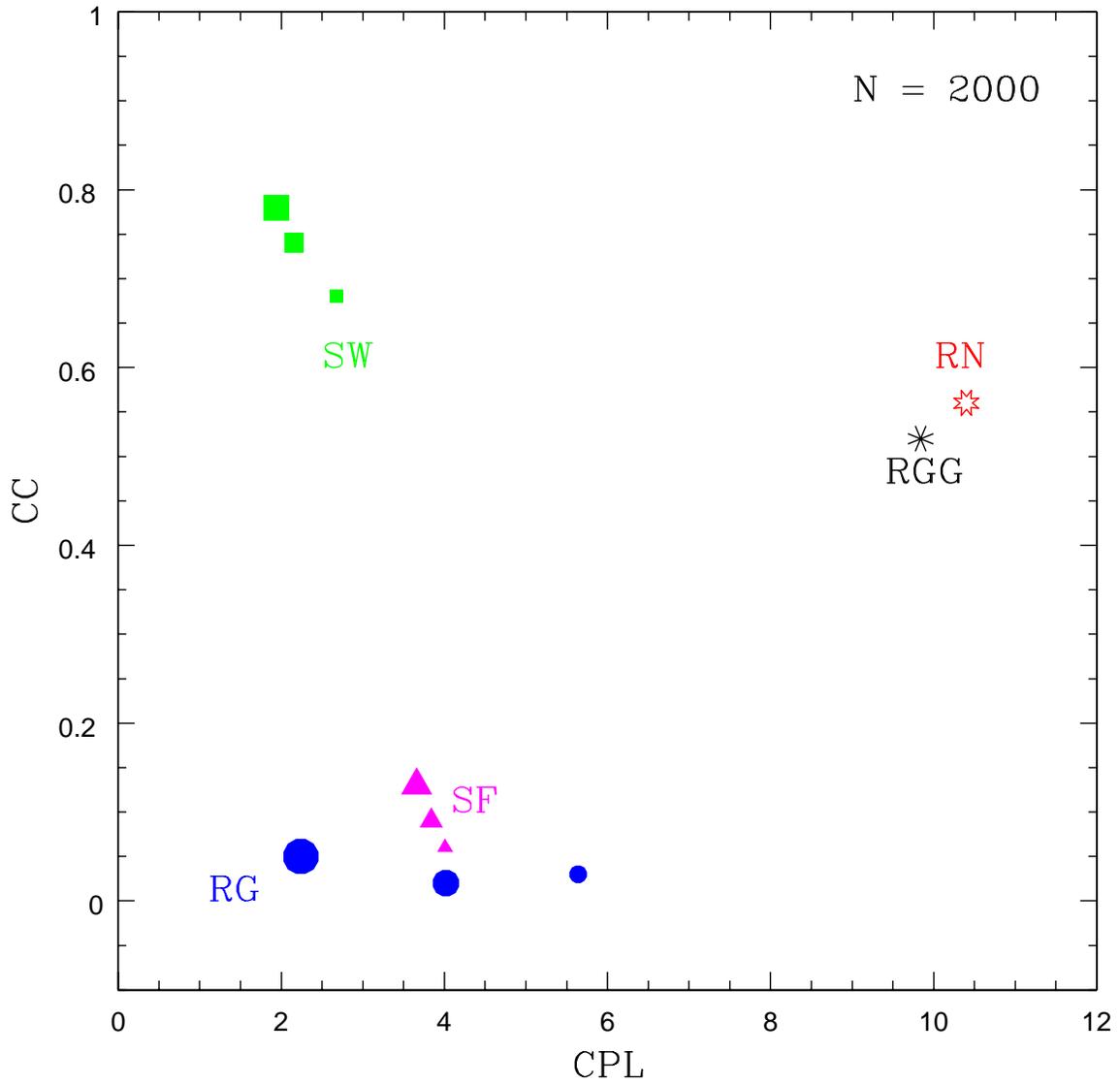}
\end{center}
\caption{CPL-CC combined plot for different classes of networks with $2000$ nodes in all cases except the  
W-S network. We show the results for $3$ SF networks (solid triangles) with $\gamma = 2.28, 2.49, 2.67$,  
$3$ RGs (solid circles) with link probability $p = 0.002, 0.0035, 0.02$ and $3$ W-S small-world networks 
(solid squares) with $100$ nodes (to reduce computation time and memory) using $\beta = 0.04, 0.05, 0.06$. 
In all cases, the size of the symbols  
increases with the values of $\gamma$, $p$ and $\beta$ respectively. The open star on the right side 
corresponds to the RN from a random time series and the asterisk corresponds to a RGG with the range 
of connection limited to that of the RNs in $3$ dimension, namely, $\epsilon = 0.1$.    
These two networks are completely different from the other networks in terms of CPL-CC values.}
\label{f.3}
\end{figure}

As the dynamical system evolves, the 
probability density over the attractor is preserved. The degree $k_i$ of a reference node $\imath$ 
represents the local connectivity of the RN and corresponds to the local phase space density around the 
reference point in the attractor. It has already been shown \cite {don1,jacob} that the local probability 
density variation is reflected in the degree distribution of the RN:  
\begin{equation}
{{k_i} \over {N}} \propto p(\vec r_i)  
 \label{eq:1}
\end{equation}
for a given $M$, where $p(\vec r_i)$ is the invariant density around the  point $\vec r_i$.
In the light of this equation, $k_i/N$  can be called the 
\emph {degree density} \cite {don3}. The above equation tells us that the distribution in terms of 
the degree density remains invariant for a given attractor as has been explicitly shown by us 
earlier \cite {jacob}. 

Due to this close connection with the invariant density, the degree distribution also carries important 
structural information on the corresponding chaotic attractor in a subtle manner. To show this explicitly, 
we consider the cumulative degree distribution of the RNs from several low-dimensional chaotic attractors. 
This representation is often used in the analysis of power-law degree distributions   
to retrieve the \emph {scaling} that is not visible in the power-law tail. It has been 
extensively discussed in the literature \cite {new1} in connection with the analysis 
of SF networks and is defined as 
\begin{equation}
F(k) = \sum_{k^{'}=k}^{k_{max}} P(k^{'})
  \label{eq:2}
\end{equation}
It can be shown that if $P(k)$  is a power-law 
with exponent $\gamma$, then $F(k)$ will also follow a power-law with 
exponent $\gamma - 1$. Here we make use of $F(k)$ only to get a better comparison 
between the degree distributions of various chaotic attractors. 

In Fig.~\ref{f.2}, we show the variation of $log$ $F(k)$ as a function of $log$ $k$ for RN from several 
low-dimensional chaotic attractors and random time series. The  parameters used for the generation of 
all the chaotic attractors are given in the Appendix of \cite {spr}. In all cases, we use the 
$y$ - component of the system with time step $0.05$ for continuous systems. Note that the position of 
the tail of the distribution follows a pattern in accordance with the dimension $D_2$ of the attractor. 
To get a possible explanation for this observation, 
we have studied the actual distribution $P(k)$ of the RNs. In all cases, 
there is a  range of $k$ values and a particular $k$ value, say $k^{\ast}$, at which 
$P(k)$ has a maximum value. The cumulative distribution starts decreasing sharply as $k$ exeeds $k^{\ast}$. 
Now, the value of $k^{\ast}$ increases with the range of $k$ values which, in turn, depends on the 
clustering of points on the attractor, as the attractor is converted into the network.  
For H\' enon and Lozi attractors, the clustering is maximum as they 
are characterized by a lower $D_2$ value and consequently the range of $k$ values is maximum. This makes  
their profile getting shifted to extreme right. As $D_2$ increases, the range of $k$ values and hence 
$k^{\ast}$ decreases due to the rescaling of all systems to the unit cube, shifting the tail to  
lower $k$ values. For the random time series, when computed from a finite time series with embedding 
dimension $M$, we get $D_2 = M$ and the range is minimum with 
$k^{\ast} = <k>$ and hence the tail of $F(k)$ is located at the extreme left. 
The above argument is supported by numerical results from earlier studies \cite {don1}, if the local 
clustering coefficient can be understood analytically from the theory of random geometric graphs as 
being associated to a ``mean local clustering dimension'' defined in \cite {don3}, which should behave 
similar to the classical correlation dimension $D_2$ considered here. 

Next, we consider how the RNs are different with respect to the other two measures, CC and CPL. We generate 
an ensemble of SF networks with different $\gamma$ values using the specific scheme of preferential 
attachment as mentioned above and RGs with different link probability $p$. 
The CPL is defined by the equation
\begin{equation}
<l> = {{1} \over {N(N-1)}} \sum_{i,j}^N l_{ij}
  \label{eq:3}
\end{equation}
where $l_{ij}$ is the shortest path length for all pairs of nodes $(\imath,\jmath)$ in the network. 
The CC of the network is defined through a local clustering coefficient $c_v$ which measures the 
probability that two randomly chosen neighbours of a given vertex $v$ are mutually linked. For 
finite graphs, this is estimated \cite {don3} in terms of relative frequency of links between the 
neighbouring vertices for the node $v$. 
\begin{equation}
c_v = {{\sum_{i,j} A_{vi}A_{ij}A_{jv}} \over {k_v(k_v - 1)}}
  \label{eq:4}
\end{equation}  
The average value of $c_v$ is taken as the CC of the whole network:
\begin{equation}
CC = {{1} \over {N}} \sum_{v} c_v
  \label{eq:5}
\end{equation}   

\begin{figure}
\begin{center}
\includegraphics*[width=16cm]{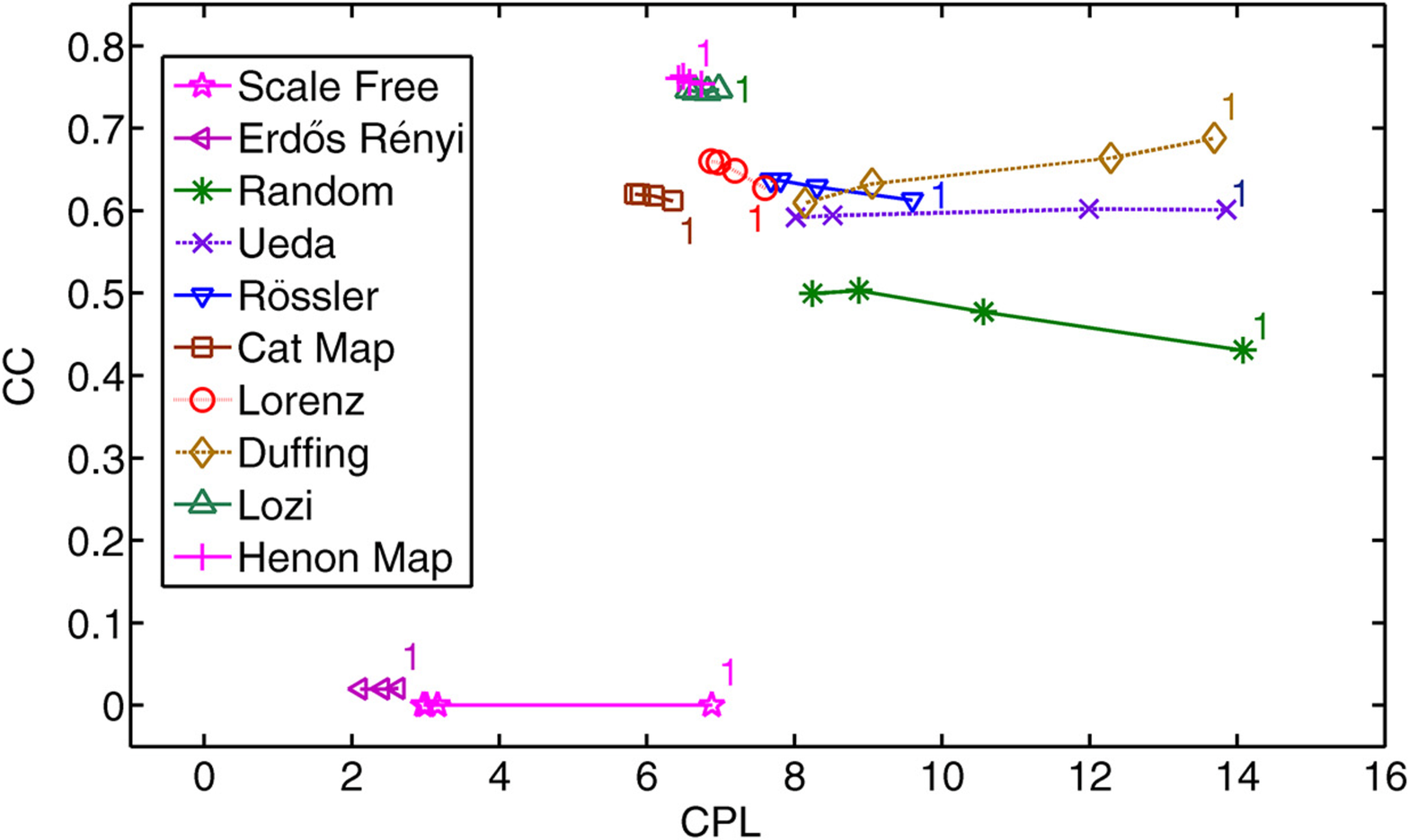}
\end{center}
\caption{CPL-CC values for RNs generated from several low-dimensional chaotic attractors and 
random time series for $4$ different values of $N$, namely, $1000, 2000, 5000$ and $10000$ with 
the point denoted as $1$ representing $N = 1000$. For comparison, values for a RG (with $p = 0.004$) 
and a SF network (with $\gamma = 2.162$) with the same $N$ values are also shown.}
\label{f.4}
\end{figure}

\begin{figure}
\begin{center}
\includegraphics*[width=16cm]{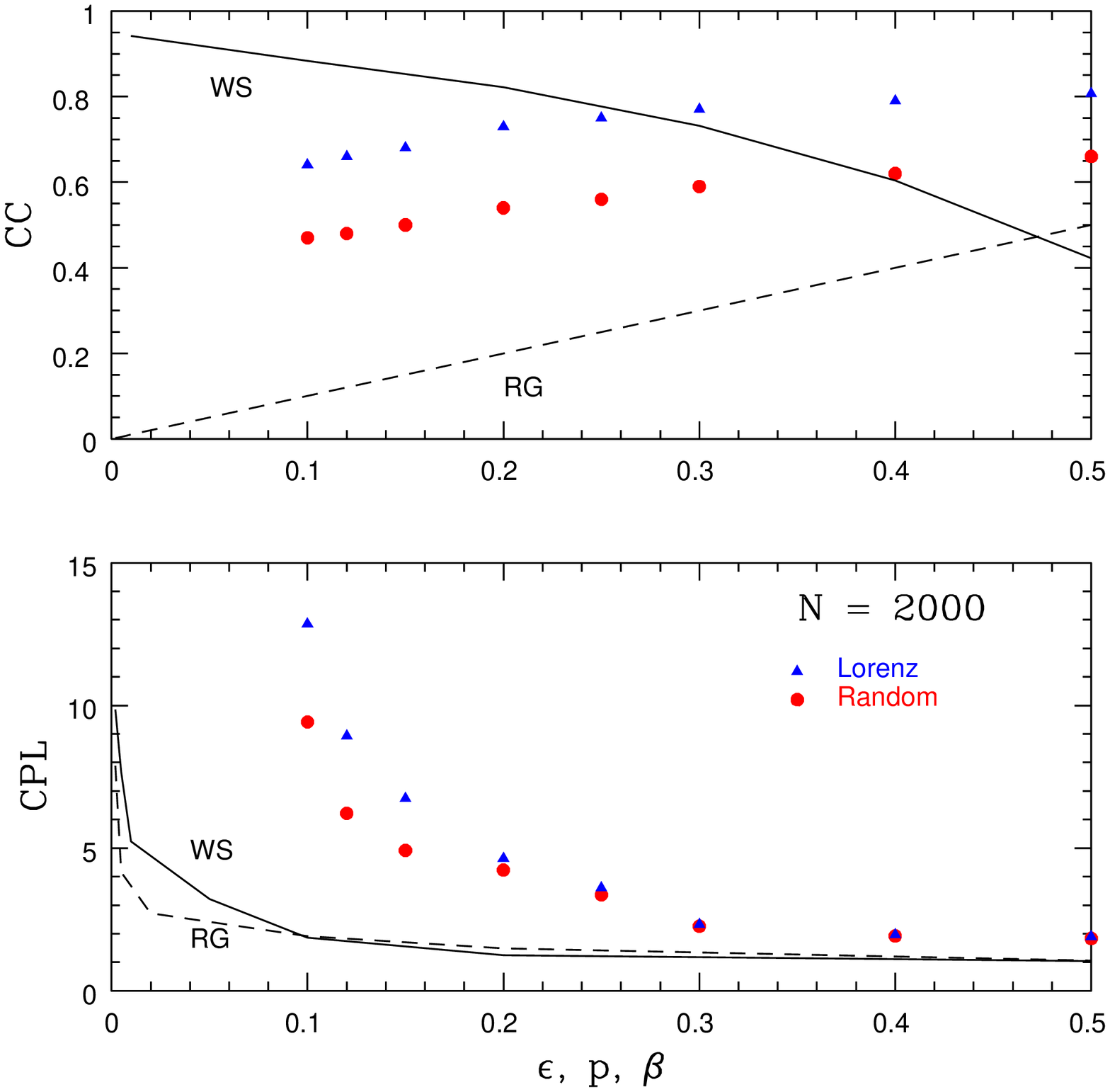}
\end{center}
\caption{Variation of the CPL and CC of RNs when the recurrence threshold $\epsilon$ is 
increased from a value close to the percolation threshold $\epsilon = 0.1$ with $M = 3$. 
The RN first undergoes a smooth transition to 
a network with high CC ($\sim 0.8$) and small CPL ($< 5$), capable of showing SW 
property for an intermediate range of $\epsilon$ approximately between $0.25$ and $0.5$.  
As $\epsilon \rightarrow 1$, the RN is found to cross over to the classical RG 
(in the limit $p \rightarrow 1)$, where both CC and CPL $\rightarrow 1$ independent of $N$. 
We also show the variation of CC and CPL for a W-S network as a function of the re-wiring probability 
$\beta$ (solid line) and the same for a RG with connection probability $p$ (dashed line). The number 
of nodes in all cases is $2000$.}
\label{f.5}
\end{figure}

The results of our computations are shown in Fig.~\ref{f.3} using a combined CPL-CC graph for $N = 2000$ in 
all cases except the W-S network. We have shown the values for the RN of a random time series and those  
from $3$ different SF networks, RGs and W-S small-world networks. The values for a RGG in $3$ dimensions  
with the range limited to that of the RN (in $3$ dimensions) are also shown. Two results are evident from the 
figure. The CPL values of all networks except the RN and RGG are very small and always $< 6$. 
Secondly, the RG with 
$p = 0.0035$, whose degree distribution coincides with that of the RN from the random time series 
behaves completely differently in 
the CPL-CC graph. The reason for this discrepancy is the absence of long range connections in the RN. 
In fact, this is true for all RNs and this property cannot change with the increase in the number of 
nodes $N$. Both CPL and CC are found to converge and saturate to a finite value for all RNs and 
cannot fall below a certain limiting value which depends on $\epsilon$. This limiting  
value is much higher compared to the typical values for SF networks and RGs. This is shown in 
Fig.~\ref{f.4} using RNs constructed from several standard chaotic attractors with $N$ varying from 
$1000$ to $10000$. Our numerical results show that RNs, constructed with $\epsilon$ within the usual 
operational window, cannot show SW property.

\begin{figure}
\begin{center}
\includegraphics*[width=16cm]{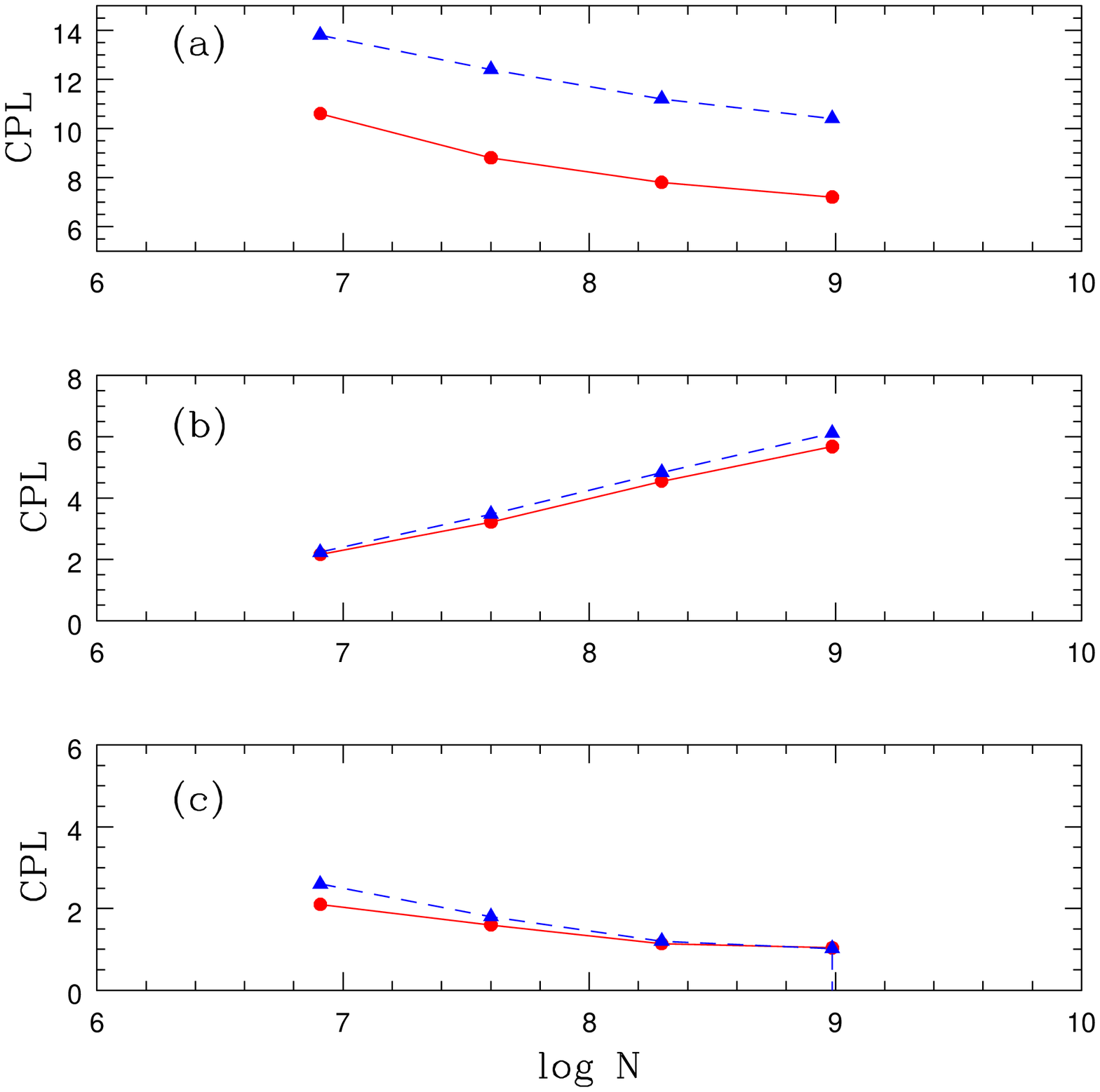}
\end{center}
\caption{Variation of CPL as a function of $log$ $N$ for RNs corresponding to the $3$ ranges of 
threshold $\epsilon$ used for the construction of the network. The top panel (a) corresponds to 
$\epsilon$ within the range of recurrence thresholds where the RN represents the characteristic 
properties of the attractor 
and the CPL converges to a limiting value as $N$ increases. The middle panel (b) corresponds to 
intermediate values of $\epsilon$ (between $0.25$ and $0.5$) where the RN is capable of displaying 
the SW property as CPL increases linearly with $log$ $N$. The bottom panel (c) corresponds to large 
values of $\epsilon$ ($> 0.5$) where the RN has long range connections and the CPL tends to a 
small value independent of $N$. For (b), the value of $\epsilon$ is tuned within the 
respective range keeping $<k>$ constant as $N$ is increased. We find the SW property for a range of 
$<k>$ values between $600$ to $1500$ depending on the value of $\epsilon$. The graph in (b) is 
plotted with $\epsilon = 0.3$ and decreasing $\epsilon$ as $N$ is increased by fixing $<k> = 800$. 
In all three cases shown above, the solid triangles 
connected by a dashed line represent the CPL of Lorenz attractor while the solid circles  
connected by a solid line represent that from random time series.}
\label{f.6}
\end{figure}

Suppose that we now 
increase the value of $\epsilon$ above the recurrence threshold which, in turn, increases the range of 
spatial connection for each node in the RN. The resulting RN may not be able to characterize the 
statistical properties of the 
embedded attractor, but one can look at the properties of the resulting network from a complex network perspective. 
For example, in the case of spatial networks, such as, communication and transportation networks, 
Barthelemy \cite {bart1,bart2} has shown that as the interaction range between nodes is of the order 
of system size or larger, the spatial effects become negligible and the networks tend to show 
SF property. RNs are \emph {spatially constrained} networks and hence we now check whether, under 
some limiting conditions, the RNs can show properties of either SW networks or classical RGs. 
  
In Fig.~\ref{f.5}, we show the variation of CC and CPL for the RNs from random time series and 
Lorenz attractor as a 
function of $\epsilon$. While the CC shows a steady increase with $\epsilon$, the CPL decreases 
much more quickly initially and saturates to the minimum possible value of $1$. We also show the CC and 
CPL for the RG (dashed line) for increasing values of $p$ and a typical Watts-Strogatz SW 
network (solid line) as a function of increasing re-wiring probability $\beta$. We find that for 
an intermediate range of $\epsilon$ values, approximately from $0.25$ to $0.5$, the RNs show 
high CC ($\sim 0.8$) and  small CPL ($< 5$) 
which are the basic criteria for a network to show the small-world property.  
To check this, we increase the value of $N$   
keeping the average degree $<k>$ approximately constant by changing the value of $\epsilon$ within the 
above range ( between $0.25$ and $0.5$) and compute the value of CPL as function of 
$log$ $N$. The result, presented in Fig.~\ref{f.6} (b), clearly shows the small-world property.   
Thus, the intermediate range of $\epsilon$ can be 
considered as a small-world phase for the RNs. As $\epsilon$ 
further increases and tends to the system size, all RNs smoothly cross over to the classical RGs with 
high $p$ where the CPL and CC $\rightarrow 1$  independent of $N$. To get a comparison of the behaviour 
between different regimes of $\epsilon$, we also show the variation of CPL with $log$ $N$ for RN 
with $\epsilon$ equal to the recurrence threshold in Fig.~\ref{f.6} (a) 
and for the case of long range connections with $\epsilon > 0.5$ in Fig.~\ref{f.6} (c). 

We have checked this result for two more standard chaotic attractors, namely, the R\"ossler attractor and 
the Duffing attractor and found an identical behavior. Hence, based on our numerical results, we 
conjecture that all RNs pass through an intermediate small-world phase as the value of 
$\epsilon$ is increased, before undergoing a transition to the classical RG for large $\epsilon$. 
Our result provides a method for constructing a SW network from  time series of dynamical systems 
by adjusting the range of connections between the nodes. The result is also relevant to search for the 
SW property in network models that require a multi-dimensional system with a metric, as for example 
in the case of spreading of diseases \cite {past}, in continuous percolation theory \cite {alon} and 
in the study of random walks on fractals \cite {jund}. For example, it is well known \cite {dall} that 
the spreading of diseases can be modeled by a RGG. If the nature of spreading is limited to 
nearest neighbour interaction (say, controlled by contact only), the network cannot show SW property. 
However, if the spreading occurs by other means, so that the range $\epsilon$ is large, the network 
can very well switch over to a SW.  

\section{Conclusion}
In this Letter, we analyse the RNs, which are unweighted and undirected graphs, from a complex network 
perspective. Three primary characteristics of a network, namely, the degree distribution, the 
CC (an averaged local measure) and the CPL (a global measure) 
are used for the analysis. We find that the degree distribution of every RN is unique and 
characteristic to the probability density variations over the representative attractor.  
By computing the CC  and 
the CPL  of several RNs from standard chaotic attractors, we explicitly show that the 
RNs constructed with the  recurrence threshold at and just above the percolation threshold are 
basically different with both CC and CPL being  
high due to the absence of long range connections between the nodes. 

The characteristic properties of the RN change with the value of $\epsilon$ used to construct it. 
As $\epsilon \rightarrow 0$, CPL $\rightarrow \infty$ and CC $\rightarrow 0$. For values of 
$\epsilon$ below the recurrence threshold, the RN consists of several disconnected clusters. 
For $\epsilon$ within the operational window above the percolation threshold,  
the network becomes globally connected (with the appearance of a giant component) and the RN measures 
characterize the statistical properties of the underlying system. In this phase, CPL 
remains relatively high (with the network unable show SW property) and converges to an asymptotic 
value as $N$ is increased.  

When the value of $\epsilon$ is increased further,  the constraints induced by the short connectivity range  
slowly disappear with the emergence of long range connections and 
all the RNs initially cross over to a SW phase with high CC and small CPL that increases as $log$ $N$. 
Finally,  as the value of 
$\epsilon \rightarrow 1$, the RN undergoes a smooth transition to classical RGs with $p \sim 1$ with 
both CPL and CC $\rightarrow 1$ independent of the number of nodes N. Thus, for all RNs, there are 
basically three  
phases in terms of the range of $\epsilon$ values used for the construction of the RN, namely,  
a small operational window of $\epsilon$ that captures the statistical properties of the 
corresponding attractor, an intermediate range where the RN can display the SW property and 
finally $\epsilon$ of the order of the size of the attractor where the RN crosses over to the 
classical RG.

{\bf Acknowledgements} 

We thank one of the anonymous referees for several suggestions to improve our manuscript. 

RJ and KPH acknowledge the financial support from Science and Engineering Research Board (SERB), 
Govt. of India in the form of a Research Project No SR/S2/HEP-27/2012. 
KPH  acknowledges the computing facilities in IUCAA, Pune.  

For graphical representation of networks, we use the GEPHI software: \\  
\emph {(https://gephi.org/)}.

\bibliographystyle{elsarticle-num}

\begin{thebibliography}{00}

\bibitem{str1}
S. H. Strogatz, ``Exploring complex networks'',  
\textit{Nature} {\bf 410}, (2001) 268--76.

\bibitem{new1}
M. E. J. Newman, \textit{Networks: An introduction}, (Oxford University Press, New York, 2010). 

\bibitem{bar1}
B. Barzel and A. L. Barabasi, ``Universality in network dynamics'',  
\textit{Nature Phys.} {\bf 9}, (2013) 673--81.

\bibitem{wat1}
D. J. Watts, \textit{Six Degrees: The Science of a Connected Age}, 
(Norton, New York, 2003).

\bibitem{was}
S. Wasserman and K. Faust, \textit{Social Network Analysis: Methods and Applications}, 
(Cambridge University Press, Cambridge, 1994).

\bibitem{new2}
M. E. J. Newman, ``The structure and function of complex networks'',  
\textit{SIAM Rev.} {\bf 45}, (2003) 167--256.

\bibitem{law}
S. Lawrence and C. L. Giles, ``Accessibility of information on the web'',  
\textit{Nature} {\bf 400}, (1999) 107--10.

\bibitem{bul}
E. Bullmore and O. Sporns, ``Complex brain networks: graph theoretical analysis of structural and 
functional systems'',  \textit{Nat. Rev. Neurosci.} {\bf 10}, (2009) 187-98.

\bibitem{lon}
P. Long, D. B. Liu, S. M. Cai, L. Hong and P. L. Zhou, 
``Recurrence network analysis of the synchronous EEG time series in normal'',  
\textit{Cell Biochem. Biophys.} {\bf 66}, (2013) 331--38.

\bibitem{erd}
P. Erd\H os and A. R\' enyi, ``On the evolution of random graphs'',  
\textit{Publ. Math. Inst. Hung. Acad. Sci.} {\bf 5}, (1960) 17--61.

\bibitem{bar2}
A. L. Barabasi and R. Albert, ``Emergence of scaling in random networks'',  
\textit{Science} {\bf 286}, (1999) 509--12.

\bibitem{alb1}
R. Albert and A. L. Barabasi, 
``Topology of evolving networks: local events and universality'',  
\textit{Phys. Rev. Lett.} {\bf 85}, (2000) 5234--37.

\bibitem{fal}
M. Faloutsos, P. Faloutsos and C. Faloutsos, 
``On power-law relationships of the internet topology'',  
\textit{Comput. Commun. Rev.} {\bf 29}, (1999) 251--62.

\bibitem{gir}
M. Girvan, and M. E. J. Newman, ``Community structure in social and biological networks'',  
\textit{Proc. Nat. Acad. Sci. USA} {\bf 99}, (2002) 7821--26.

\bibitem{gui}
R. Guimera and L. A. N. Amaral, ``Functional cartography of complex metabolic networks'',  
\textit{Nature} {\bf 433}, (2005) 895--900.

\bibitem{alb2}
R. Albert, I. Albert and G. L. Nakarado, 
``Structural vulnerability of the North American power grid'',  
\textit{Phys. Rev. E} {\bf 69}, (2004) 025103(R).

\bibitem{wat2}
D. J. Watts and S. H. Strogatz, ``Collective dynamics of small world networks'',  
\textit{Nature} {\bf 393}, (1998) 440--42.

\bibitem{kan}
H. Kantz and T. Schreiber, \textit{Nonlinear Time Series Analysis}, 
(Cambridge University Press, Cambridge, 2004).

\bibitem{hil}
R. C. Hilborn, \textit{Chaos and Nonlinear Dynamics}, 
(Oxford University Press, Oxford, 1994). 

\bibitem{zha}
J. Zhang and M. Small, 
``Complex networks from pseudoperiodic time series:topology versus dynamics'',  
\textit{Phys. Rev. Lett.} {\bf 96}, (2006) 238701.

\bibitem{lac}
L. Lacasa, B. Luque, F. Ballesteros, J. Luque and J. C. Nuno, 
``From time series to complex networks: the visibility graph'',  
\textit{Proc. Nat. Acad. Sci. USA} {\bf 105}, (2008) 4972--75. 

\bibitem{mar1}
N. Marwan, J. F. Donges, Y. Zou, R. V. Donner and J. Kurths, 
``Complex network approach for recurrence analysis of time series'',  
\textit{Phys. Lett. A} {\bf 373}, (2009) 4246--54.

\bibitem{eck}
J. P. Eckmann, S. O. Kamphorst and D. Ruelle, 
``Recurrence plot of dynamical systems'',  
\textit{Europhys. Letters} {\bf 5}, (1987) 973--77.

\bibitem{gra1}
P. Grassberger and I. Procaccia, ``Measuring the strangeness of strange attractors'',  
\textit{Physica D} {\bf 9}, (1983) 189--208.

\bibitem{kxu}
X. K. Xu, J. Zhang and M. Small, 
``Superfamily phenomena and motifs of networks induced from time series'', 
\textit{Proc. Nat. Acad. Sci. USA} {\bf 105}, (2008) 19601--05.

\bibitem{don1}
R. V. Donner, Y. Zou, J. F. Donges, N. Marwan and J. Kurths, 
``Recurrence networks: A novel paradigm for nonlinear time series analysis'', 
\textit{New J. Phys.} {\bf 12}, (2010) 033025.

\bibitem{don2}
R. V. Donner, M. Small, J. F. Donges, N. Marwan, Y. Zou, R. Xiang and J. Kurths, 
``Recurrence based time series analysis by means of complex network methods'', 
\textit{Int. J. Bif. Chaos} {\bf 21}, (2011) 1019--46.

\bibitem{mar2}
N. Marwan and J. Kurths, 
``Complex network based techniques to identify extreme events and (sudden) transitions in spatio-temporal systems'',  
\textit{CHAOS} {\bf 25}, (2015) 097609.

\bibitem{avi}
R. Avila, A. Gapelyuk, N. Marwan, T. Walther, H. Stepan, J. Kurths and N. Wessel,  
``Classification of cardio-vascular time series based on different coupling structures using recurrence network analysis'', 
\textit{Philos. T. Roy. Soc. A} {\bf 371}, (2013) 20110623.

\bibitem{zou}
Y. Zou, J. Heitzig, R. V. Donner, J. F. Donges, J. D. Farmer, R. Meucci, S. Euzzor, N. Marwan and 
J. Kurths, ``Power laws in recurrence networks in dynamical systems'', 
\textit{Europhys. Letters} {\bf 98}, (2012) 48001--06.

\bibitem{don3}
R. V. Donner, J. Heitzig, J. F. Donges, Y. Zou, N. Marwan and J. Kurths, 
``The geometry of chaotic dynamics-A complex network perspective'', 
\textit{European Phys. J. B} {\bf 84}, (2011) 653--72.

\bibitem{dong}
J. F. Donges, J. Heitzig, R. V. Donner and J. Kurths, 
``Analytical framework for recurrence network analysis of time series'', 
\textit{Phys. Rev. E} {\bf 85}, (2012) 046105.

\bibitem{dall}
J. Dall and M. Christensen, ``Random geometric graphs'',  
\textit{Phys. Rev. E} {\bf 66}, (2002) 016121.

\bibitem{rava}
A. L. Barabasi, E. Ravasz and T. Vicsek, 
``Deterministic scale-free networks'', 
\textit{Physica A} {\bf 299}, (2001) 559--564.
 
\bibitem{kph1}
K. P. Harikrishnan, R. Misra, G. Ambika and A. K. Kembhavi, 
``A non-subjective approach to the GP algorithm for analysing noisy time series'', 
\textit{Physica D} {\bf 215}, (2006) 137--145.

\bibitem{kph2}
K. P. Harikrishnan, R. Misra and G. Ambika, 
``Revisiting the box counting algorithm for the correlation dimension analysis of hyperchaotic time series'', 
\textit{Comm. Nonlinear Sci. Num. Simulations} {\bf 17}, (2012) 263--276.

\bibitem{kphk11}
Fortran codes for computing $D_2$ and $K_2$ from time series can be freely downloaded from the 
website: $https://sites.google.com/site/kphk11/home$

\bibitem{jacob}
R. Jacob, K. P. Harikrishnan, R. Misra and G. Ambika, 
``Uniform framework for the recurrence-network analysis of chaotic time series'', 
\textit{Phys. Rev. E} {\bf 93}, (2016) 012202.

\bibitem{spr}
J. C. Sprott, \textit{Chaos and Time Series Analysis}, 
(Oxford University Press, Oxford, 2003).

\bibitem{bart1}
M. Barthelemy, ``Spatial Networks'',  \textit{Phys. Reports} {\bf 499} (2011) 1--101.

\bibitem{bart2}
M. Barthelemy, ``Crossover from scale free to spatial networks'', 
\textit{Europhys. Letters} {\bf 63}, (2003) 915--18.

\bibitem{past}
R. Pastor-Satorras and A. Vespigniani, ``Epidemic spreading in scale-free networks'', 
\textit{Phys. Rev. Lett.} {\bf 86}, (2001) 3200--03.

\bibitem{alon}
U. Alon, A. Drory and T. Balbery, 
``Systematic derivation of percolation thresholds in continuum systems'', 
\textit{Phys. Rev. A} {\bf 42}, (1990) 4634--38.

\bibitem{jund}
P. Jund, R. Jullien and T. Campbell, 
``Random walks on fractals and stretched exponential relaxation'', 
\textit{Phys. Rev. E} {\bf 63}, (2001) 036131.



\end{thebibliography}

\end{document}